\documentclass[pra,aps,reprint,showpacs,amsmath,superscriptaddress]{revtex4-1}

\usepackage{bm}
\usepackage{graphicx}
\usepackage{dsfont}

\usepackage{mathrsfs}

\usepackage{framed}
\usepackage{multirow}
\usepackage[all]{xy}

\usepackage{framed}
\usepackage{comment}
\usepackage{here}                    
\usepackage{latexsym}		     
\usepackage{amsmath}     
\usepackage{amsfonts}  
\usepackage{amssymb}
\usepackage{amsthm}
\usepackage{dcolumn}
\usepackage{color}
\usepackage{mathrsfs}
\usepackage{dsfont}

\renewcommand{\phi}{\varphi}

\newcommand{\be}{\begin{equation}}
\newcommand{\ee}{\end{equation}}
\newcommand{\bea}{\begin{eqnarray}}
\newcommand{\eea}{\end{eqnarray}}

\renewcommand{\phi}{\varphi}

\usepackage[colorlinks,citecolor=blue,linkcolor=blue,urlcolor=blue]{hyperref}

\begin{document}

\title{Effects of Kerr nonlinearity in physical unclonable functions}

\author{Georgios M. Nikolopoulos}

\affiliation{Institute of Electronic Structure and Laser, Foundation for Research and Technology-Hellas, GR-70013 Heraklion, Greece}

\date{\today}

\begin{abstract} 
We address the question of whether the presence of Kerr  
nonlinearity in multiple-scattering optical media  offers any  advantage 
with  respect to the design of physical unclonable 
functions. Our results suggest that under certain conditions, 
nonlinear physical unclonable functions can be more robust against the 
potential cloning of the medium, relative to their linear counterparts 
that have been exploited in the context of various cryptographic applications. 
\end{abstract}

\maketitle


\section{Introduction}
 
Physical unclonable functions (PUFs) based on optical multiple scattering media have attracted considerable attention over the last decades, mainly because they are considered to 
pave the way for the development of novel more robust cryptographic protocols 
\cite{Pappu02,Goorden14,Uppu19,Iesl16,NikDiaSciRep17,Nik18,Mes18,FlaNikAlbFis19,Nik21,GiaKamBru20,Wang-etal21, Horstmayer13,Buch05,Yeh12,Nik19,Chow-etal21}. 
The optical multiple scattering medium essentially serves as a token, 
and the internal disorder renders its cloning a formidable technological challenge.
At the same time,  the  response of the  token to optical challenges 
is a complex pattern (speckle),  which depends strongly on the internal disorder of 
the token as well as on the challenge. Different challenges pertain to different parameters of the incoming light, e.g., intensity, point and angle of incidence, wavelength or combinations thereof \cite{Pappu02,Iesl16,Mes18,Nik21, Horstmayer13}. Formally speaking, 
the optical response of the token involves a large number of optical modes, and the speckle is essentially the result of interference of many paths that lead to a particular mode at the output \cite{Goodman1}. 
In the framework of  cryptographic applications, the  speckle can be used raw as a fingerprint \cite{Iesl16,Ruh-etal}, or it 
can be processed classically to yield a random numerical key  \cite{Pappu02,Mes18}.  In any case, the sensitivity of the 
speckle to the potential cloning of the token, essentially determines 
the resistance of the cryptographic protocol under consideration against related attacks.
A useful protocol should allow one to distinguish between a token and its clones, by comparing their responses to the same challenge (raw speckles or numerical keys).

All of the aforementioned studies rely on the use of a linear multiple scattering medium as token, which results in a linear relationship between the recorded intensity at a particular point at the output plane, and the electric field at the input plane. This linear relationship may render certain cryptographic protocols vulnerable to cloning attacks, provided the attacker has access to the raw speckle images that are produced by the setup for the particular token. It has been conjectured 
\cite{Pappu02,Ruh-etal} that such attacks can be thwarted by using nonlinear multiple scattering tokens, thereby breaking the linear dependence of the speckle on the electric field at the input. To the best of our knowledge, this conjecture remains an open question in the field up to date. 

In this work, our aim is to address for the first time this open question, by 
investigating  whether and under what conditions multiple scattering media with Kerr nonlinearity can offer some advantage over their linear counterparts, with respect to their robustness against cloning. To this end we focus on the speckle  that is produced by a reference token, and we study its 
correlation to the speckles of different random 
clone tokens. Given the high sensitivity of the speckle to the internal disorder, we adopt a generic model for the (non)linear token, which allows us to reach definite general conclusions through the statistical analysis of the data. Our results suggest that, in general, speckles produced by tokens with focusing nonlinearity can be more robust to cloning, than speckles produced by linear tokens, or tokens with defocusing nonlinearity.  


\section{Formalism} 

In the diffusive limit \cite{NikDiaSciRep17,Nik18,FlaNikAlbFis19,Goodman1}, the problem can be formulated in terms of ${\mathscr M}$ input and ${\mathscr M}$ output transverse spatial modes. For monochromatic laser light at wavelength $\lambda$, the electric field at the $m$th output mode is given by \cite{AndVolPopKatGreGig15,MouAndDefVolKatGreGig16,BeiPutLagMos11,SmaKatGuaSil12}
\begin{equation}
E_m^{\rm (out)}(\lambda) = \sum_{j=1}^{\mathscr M} T_{m,j}^{(\lambda)}  E_j^{\rm (in)}(\lambda),
\label{Em:eq}
\end{equation}
where $T_{m,j}^{(\lambda)} $ is the element of the multi-spectral transmission matrix linking the $m$th output mode, to the $j$th input mode, for light at wavelength $\lambda$. In the notation adopted throughout this work, the $i$th input and output transverse modes are associated with spatial   coordinates $(x_i,y_i):={\bm \rho}_i$ at the input and output planes, 
 which are located at $z=z_1$ and $z=z_{2}$, respectively.  Hence, 
$E_m^{\rm (out)}(\lambda)$ and $E_j^{\rm (in)}(\lambda)$  are abbreviations to  $E({\bm \rho}_m,z_2;\lambda)$ and $E({\bm \rho}_j,z_1;\lambda)$, respectively.
For a given wavelength, the 
elements of the transmission matrix $\{T_{m,j}^{(\lambda)}\} $ depend strongly on the realization of the disorder in the token, and can be treated as independent complex Gaussian random variables, with zero mean and variance given by $(1-l/L)/{\mathscr M}$, where $L$ is the thickness of the token and  $l \ll L$ is the mean free path. 
For a  token with spectral correlation bandwidth 
$\delta\lambda_{\rm c}$,  
the transmission matrices at two different wavelengths $\lambda_1$ and $\lambda_2$, with $|\lambda_1-\lambda_2|> \delta\lambda_{\rm c}$, can be treated as 
independent random matrices, 
thereby yielding uncorrelated monochromatic speckle patterns 
at the two  wavelengths \cite{AndVolPopKatGreGig15,MouAndDefVolKatGreGig16,BeiPutLagMos11,SmaKatGuaSil12}.

\begin{figure}
\centering\includegraphics[width=8.5cm]{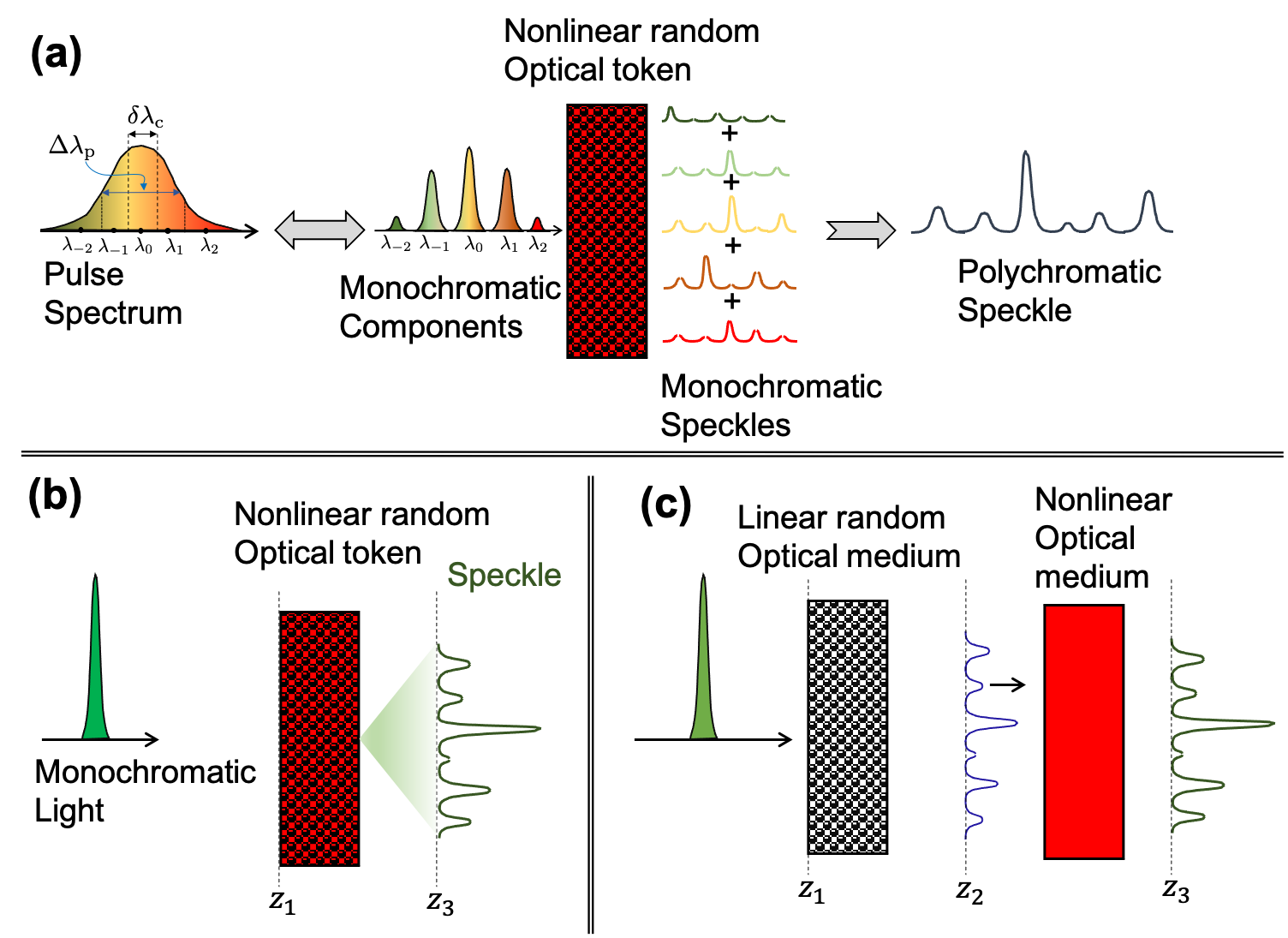}
\caption{Schematic representation of the theoretical model under consideration, for the scattering of a Fourier-limited laser  pulse from a multiple scattering nonlinear token. (a) For a token with spectral correlation bandwidth $\delta\lambda_{\rm c}$ and a pulse of bandwidth $\Delta\lambda_{\rm p}$, there 
are ${\mathscr N}\simeq \Delta\lambda_{\rm p}/\delta\lambda_{\rm c}$ spectral 
degrees of freedom, corresponding to wavelengths $\lambda_0,\lambda_{\pm 1},\ldots$. Different  
spectral components pertain to independent random monochromatic transmission matrices, and they give rise to uncorrelated monochromatic speckles. (b,c) The scattering of the $i$th spectral component from the nonlinear random token is treated as a sequential process, pertaining to the scattering from a linear multiple-scattering token, and the propagation of the resulting fully-developed monochromatic speckle in a nonlinear homogeneous medium. }
\label{figure1}
\end{figure}

Equation (\ref{Em:eq}) refers to the scattering of monochromatic light with wavelength $\lambda$, from a disordered linear token.
To exploit the nonlinearity of a medium, one typically needs 
ultrashort laser pulses, so that to reach high intensities without damaging the medium. The spectrum of an ultrashort laser pulse involves many frequencies, and the power spectral density defines the power in each wavelength of the spectrum. Neglecting the spatial dependence of the field for a moment, for a Gaussian pulse of the form 
$E(t) = \sqrt{{I}_0} \exp[-t^2/2\tau^2]\exp(-{\rm i}2\pi\nu_0 t)$,  
the electric field at frequency $c = \lambda \nu$ is given by 
\begin{equation}
E(\nu) =\sqrt{2\pi {I}_0}\tau 
\exp\left [-2\pi^2\tau^2(\nu-\nu_0)^2\right ]. 
\label{Elambda:eq}
\end{equation} 
The bandwidth (FWHM) of the pulse $|E(\nu)|^2$ is $\Delta\nu_{\rm p} = \sqrt{\ln(2)}/(\pi  \tau)$, 
${I}_0$ and $T=2\sqrt{\ln(2)}\tau$ denote the peak intensity (units of  ${\rm W}/{\rm cm}^2$) and the FWHM of the pulse in time domain. The bandwidth can be expressed in terms of the wavelength using the relation $\Delta\lambda_{\rm p} = \lambda_0^2\Delta\nu_{\rm p}/c$. 

The spectral density is $S(\nu)= |E(\nu)|^2$, and $S(\nu)d\nu$ gives the fluence (energy per area) for frequencies between $\nu-d\nu/2$ and $\nu+d\nu/2$.  
Hence, the intensity carried by frequencies within the spectral correlation bandwidth 
$\delta\nu_{\rm c}$ around $\nu$, can be approximated by  
\begin{equation}
J(\nu):=\frac{S(\nu)\delta\nu_{\rm c}}{T} =  \frac{I_0}{{\mathscr N}}
\exp\left [-4\pi^2\tau^2(\nu-\nu_0)^2\right ],
\label{Jnu:eq}
\end{equation}  
 where ${\mathscr N}\equiv \Delta\nu_{\rm p} / \delta\nu_{\rm c}$ represents 
 the spectral degrees of freedom i.e., the number of spectral channels.  When a pulse 
 of bandwidth $\Delta\nu_{\rm p} $ propagates through a multiple-scattering random token with spectral correlation bandwidth $\delta\nu_{\rm c}$, the generated 
speckle is essentially a superposition of ${\mathscr N}$ uncorrelated monochromatic speckle patters 
[see Fig. \ref{figure1}(a)], with the exponent in Eq. (\ref{Jnu:eq}) determining the weight of the different contributions \cite{AndVolPopKatGreGig15,MouAndDefVolKatGreGig16,BeiPutLagMos11,SmaKatGuaSil12}.

Let $E({\bm \rho},z_2;\lambda_i)$  denote the two-dimensional speckle field at the output plane, for the 
$i$th component pertaining to wavelengths in the interval $[\lambda_i-\delta\lambda_{\rm c}/2,\lambda_i+\delta\lambda_{\rm c}/2]$, with $\lambda_i = c/\nu_i$. 
Assuming uniform illumination,  the electric field at the $m$th output mode 
is given by 
\begin{subequations}
\label{LinearField_ith:eq}
\begin{equation}
E_m^{\rm (out)}(\lambda_i) \equiv E({\bm \rho}_m,z_2;\lambda_i) =\sqrt{J(\nu_i)}\Psi({\bm \rho}_m,z_2;\lambda_i),
\label{Em:eq2}
\end{equation}
where 
\begin{equation}
 \Psi({\bm \rho}_m,z_2;\lambda_i) = \frac{1}{\sqrt{{\mathscr M}}} 
 \sum_{j=1}^{\mathscr M} T_{m,j}^{(\lambda_i)} 
 \label{Psi:eq}
\end{equation}
encapsulates the effect of the internal random disorder of the token on the spatial profile of the electric field at the particular wavelength (see Ref. \cite{NikDiaSciRep17} and references therein). In the discretized model under consideration, 
the total speckle field at wavelength $\lambda_i$ is given by 
\begin{equation}
E({\bm \rho},z_2;\lambda_i)\equiv\{E_m^{\rm (out)}(\lambda_i): 1\leq m\leq {\mathscr M}\}, 
\label{Etot:eq}
\end{equation}
where $E_m^{\rm (out)}(\lambda_i)$ is given by Eqs. (\ref{Em:eq2}) and (\ref{Psi:eq}).
\end{subequations}

What remains to be done, is the inclusion of nonlinearity in our model for the randomly disordered  token. In general, the propagation of light in a nonlinear inhomogeneous medium is a rather 
challenging and time consuming task, because one has 
to know the details of scatterers (including position, size, and shape). 
Even if one models successfully a particular token with well-defined disorder, 
it is not clear at all what the results that will be obtained may imply for other tokens 
pertaining to different independent realizations of random disorder. 
Given that our task here is to investigate whether the nonlinearity can 
make speckle patterns more resilient to cloning, we will adopt a  generic 
model for the token, consisting of a thin linear multiple-scattering medium, which is followed by a homogeneous layer with Kerr-type nonlinearity (see Fig. \ref{figure1}). A similar model has been used successfully  
in studies related to wavefront shaping techniques in the presence of disorder and nonlinearity \cite{FroSmaDanOulDerSil17}. 

The speckle field $E({\bm \rho},z_2;\lambda_i)$ for the $i$th spectral component at the output of the linear random medium  is given by Eqs. (\ref{LinearField_ith:eq}). The propagation of this monochromatic speckle field in  the nonlinear medium of thickness $L$, will be modeled by the following nonlinear 
wave equation \cite{Boyd08}
\begin{widetext}
\begin{equation}
 {\rm i}\frac{\partial \Psi({\bm r};\lambda_i)}{\partial z} + 
 \frac{1}{2k_in_0} \nabla_\perp^2  \Psi({\bm r};\lambda_i) +
k_in_2 J(\nu_i)|\Psi({\bm r},\lambda_i)|^2 \Psi({\bm r};\lambda_i)=0, 
\label{weq1}
\end{equation}
\end{widetext}
where $\nabla_\perp^2\equiv
 \frac{\partial^2 }{\partial {x}^2} +  \frac{\partial^2 }{\partial {y}^2} 
 $, ${\bm r}\equiv ({\bm \rho}, z)$, $k_i = 2\pi/\lambda_i$, $n_0$ is the  
 linear refractive index, and the nonlinear refractive index (Kerr nonlinearity) is denoted by $n_2$ and it is measured 
in ${\rm cm}^2/{\rm W}$. This equation treats each spectral component independently, while propagation takes place on a discretized space, with 
the initial condition   
$\Psi({\bm \rho},z_2;\lambda_i) = \{\Psi({\bm \rho}_m,z_2;\lambda_i): 1\leq m\leq {\mathscr M}\}$, where  
$\Psi({\bm \rho}_m,z_2;\lambda_i)$ is given by Eq. (\ref{Psi:eq}). Note also that different spectral components experience different nonlinearities, and thus differrent dynamics, due to the intensity $J(\nu_i)$. 
Throughout this work we are interested in short propagation  lengths $z_3 = L\lesssim 1$mm, so that the token is sufficiently thin for cryptographic applications 
(e.g., engineering of all-optical smart cards) \cite{Pappu02,Nik21}. 
As a result, in Eq. (\ref{weq1}) we neglect effects of dispersion, 
self-steepening and space-time coupling, assuming pulses with duration such that 
$2\pi c T\gg \lambda_0$ \cite{Boyd08}.  
For observation times long compared to the coherence time of the scattered light $(\sim 1/\delta\nu_{\rm c})$, the intensity of the scattered light at the $m$th output mode after the nonlinear medium (plane at $z=z_3$) can be approximated by \cite{Goodman1}
\begin{equation}
I_m^{\rm (tot)} \simeq \sum_{i} J(\lambda_i)
| \Psi({\bm \rho}_m,z_3;\lambda_i) |^2. 
\end{equation}
We also neglect effects of free-space propagation, as they are expected to contribute equally in the speckles of the actual token and a clone. Without loss of generality, we consider odd values for ${\mathscr N}$, with the index $i\in [-({\mathscr N}-1)/2, ({\mathscr N}+1)/2]$, and $i=0$ refers to the central  wavelength $\lambda_0 = c/\nu_0$.

\begin{figure}
\centering\includegraphics[width=8.5cm]{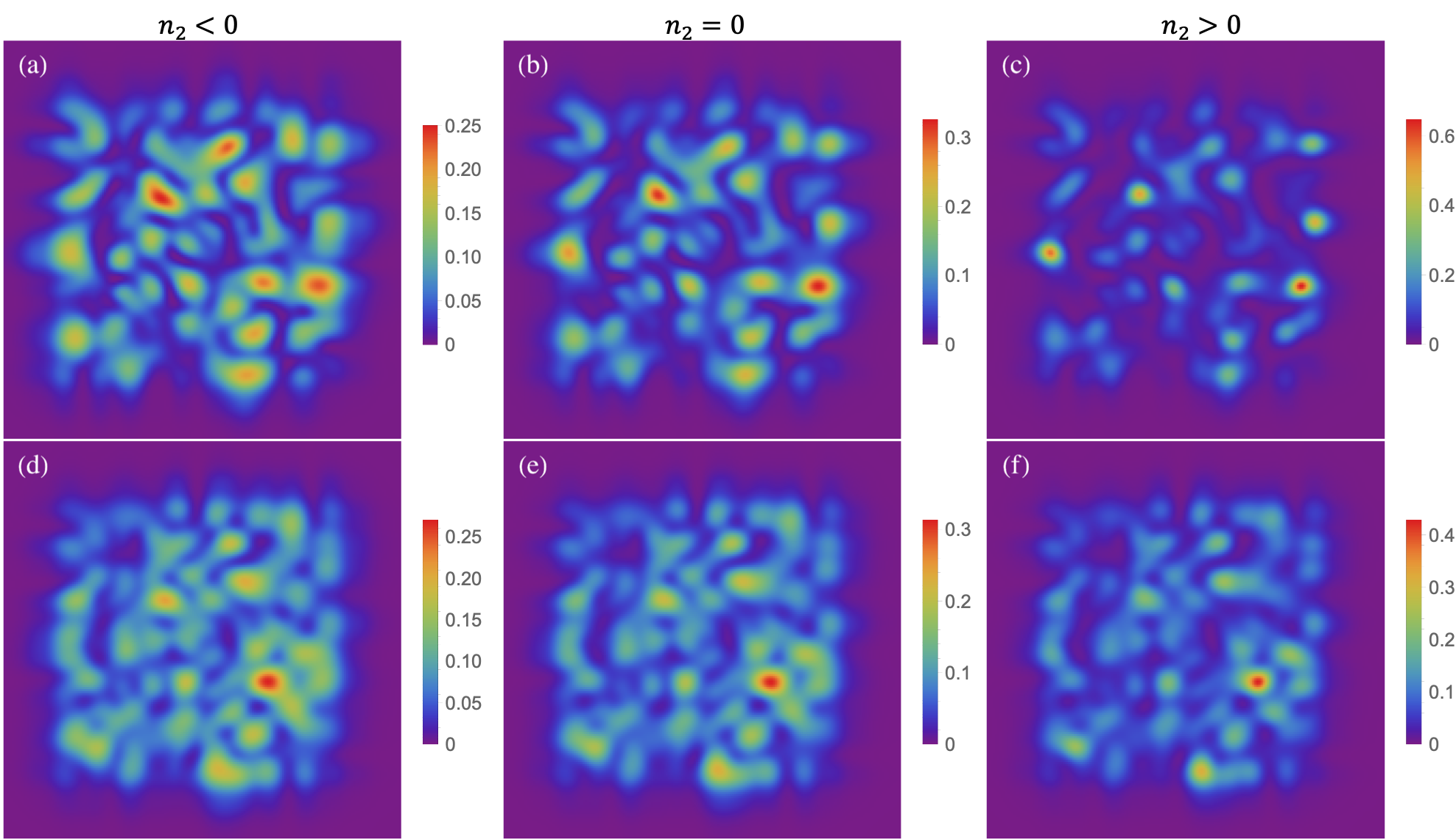}
\caption{Speckle fields after propagation in a medium with (a,d)  defocusing nonlineariry 
$n_2=-3\times 10^{-16}$cm$^2$/W, 
(b,e) zero nonlinearity, and (c,f) focusing nonlineariry $n_2=3\times 10^{-16}$cm$^2$/W. Note the different color scales as one goes from negative to positive $n_2$.  Parameters: $L = 100\lambda_0$, $\lambda_0 = 800$nm, $I_0 = 2.6\times 10^{13}$W/cm$^2$, $T = 100$fs, $n_0 = 1.47$, ${\mathscr N} = 3$ (a-c) and ${\mathscr N} = 5$ (d-f). In all of the cases, uniform illumination has been considered at the input.}
\label{figure2}
\end{figure}

Using the above sequential model for the nonlinear multiple-scattering token, we have performed a number of simulations, in order to gain insight into the role of the nonlinearity on the speckle that is produced by such a medium, as well as on the sensitivity of the speckle on the cloning of the token. Equation (\ref{weq1}) was solved by means of the split-step Fourier method \cite{Oka06}.
The reference token 
is described by a multi-spectral transmission matrix ${\bm T}_{\rm ref}:=\{T_{m,j}^{(\lambda_i)}~:~ 1\leq m,j \leq {\mathscr M};\, -({\mathscr N}-1)/2\leq i\leq ({\mathscr N}+1)/2\}$. The matrix involves $|{\bm T}_{\rm ref}| = {\mathscr M}\times {\mathscr M}\times {\mathscr N}$ independent complex Gaussian random variables,  with zero mean and variance ${\mathscr M}^{-1}(1-l/L)$ (e.g., see \cite{NikDiaSciRep17,Goodman1,AndVolPopKatGreGig15,MouAndDefVolKatGreGig16,BeiPutLagMos11,SmaKatGuaSil12} 
and references therein). The wavelength for the $i$th spectral degree of freedom is given 
by $\lambda_i = \lambda_0+i \delta\lambda_{\rm c}$. For a fixed multi-spectral transmission matrix, we  generate random clones along the lines described in Refs. \cite{NikDiaSciRep17}. 
More precisely, the  multi-spectral transmission matrix of a $Q-$close clone is obtained 
from ${\bm T}_{\rm ref}$, by choosing new random values for $\lceil Q|{\bm T}_{\rm ref}| \rceil$ randomly chosen  elements. The parameter $Q$ is essentially associated with  
the quality of the cloning, with $Q=0$ and 1, corresponding to perfect and fully randomized clone, respectively.  


\section{Results} 

Let us begin with the effect of Kerr nonlinearity on the speckle generated by a given 
token. The main findings are summarized in Fig. \ref{figure2}, which shows 
the speckles for a token with defocusing nonlineariry $(n_2<0)$,  zero nonlinearity 
$(n_2=0)$, and focusing nonlinearity $(n_2>0)$. 
For defocusing nonlineariry the scattered light is distributed among a larger number of output modes, relative to the linear case. By contrast, for 
focusing nonlinearity the scattered light tends to concentrate within a small number of output modes, thereby leading to a small number of very bright spots in the pattern 
(note the different color scales in the density plots). 
These findings were present for all of the simulations we have performed, and 
the location as well as the intensities of these bright spots depend on the realization of the random disorder in the token. Moreover, we have found that the concentration of scattered light for  focusing nonlinearity becomes  weaker for increasing number of spectral degrees of freedom. This was to be expected, because according to 
Eqs. (\ref{Jnu:eq}) and (\ref{weq1}), effects of nonlinearity become weaker as we increase ${\mathscr N}$. Analogous behavior is found as we decrease $I_0$, for fixed 
${\mathscr N}$.  

Typical PUF-based cryptographic protocols which use optical tokens exploit one way or another the strong dependence of the speckle image on the random internal disorder of the token as well as on 
the parameters of the input light \cite{Pappu02,Iesl16,Mes18,Nik21, Horstmayer13}. 
In either case, the robustness of a protocol against the cloning of the token is mainly determined by the sensitivity of the speckle image to the cloning. 
To investigate the role of the nonlinearity in this context, for a fixed reference token, 
we generated 100 independent random clones, for three different cloning factors $Q$. 
For each clone, we calculated the speckle 
and compared it to the corresponding speckle for the reference token, using 
the Pearson correlation coefficient \cite{Ross14}, which is equal to  
$1$ for perfectly correlated patterns, and equal to 0 for totally uncorrelated patterns. 
Given that the correlation varies from clone to clone, we have analyzed the recorded values 
in terms of their median, spread and skewness, through box and whisker diagrams. 

\begin{figure}
\centering\includegraphics[width=8.5cm]{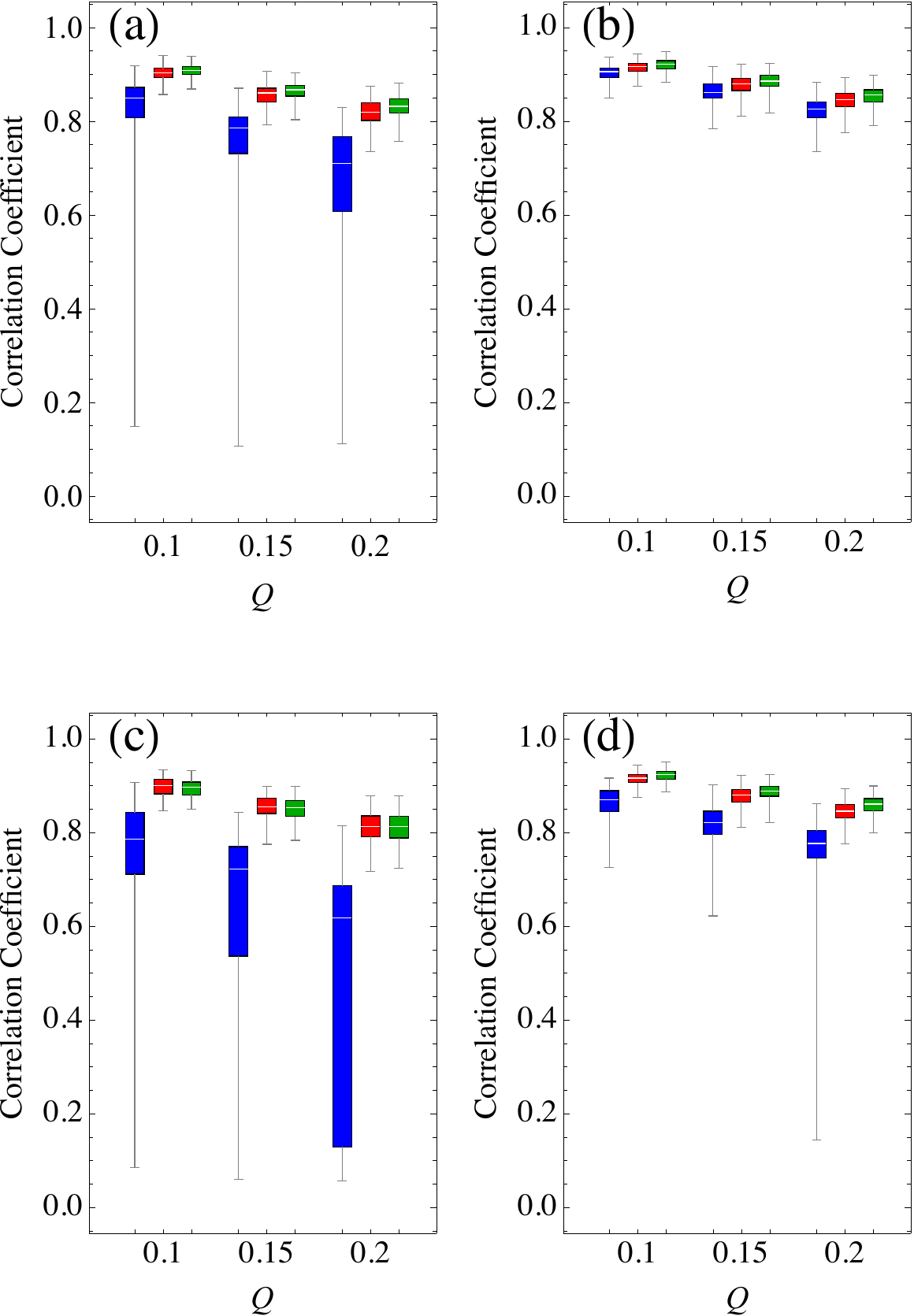}
\caption{Box and whisker diagrams for the estimated Pearson correlation coefficients 
for cloning factors $Q=(0.1,0.15,0.2)$, in the case of token with focusing nonlineariry  $n_2=3\times 10^{-16}$cm$^2$/W (blue), defocusing nonlineariry $n_2=-3\times 10^{-16}$cm$^2$/W (green) and zero nonlinearity (red). For each $Q$, 100 different random clones were generated. The lower and upper bounds of a box refer to the first and the third quartiles, and the horizontal (white line) the median. The vertical whiskers (error bars) show the minimum and the maximum of the recorded coefficients. 
(a)  $\lambda_0 = 800$nm, $n_0=1.47$, $l/L=0.2$, $T = 200$fs,  $I_0 = 1.0\times 10^{13}$W/cm$^2$, $L = 200\lambda_0$,  ${\mathscr N} = 1$ slice.
(b) As in (a) but for ${\mathscr N} = 3$ slices. (c)  As in (a) but for $L = 400\lambda_0$. 
(d) As in (b) but for $I_0 = 2.0\times 10^{13}$W/cm$^2$. }
\label{figure3}
\end{figure}

A sample of our main findings is depicted in Fig. \ref{figure3}, where we show diagrams for a fixed reference token, as a function of the cloning factor $Q$, for various combinations of parameters. As expected, the correlation decreases with increasing $Q$, because the quality of the clones decreases.  
Moreover, we find a large overlap between the red and green boxes for all of the parameters, 
which suggests that the performance of the linear token (red boxes) is the same as (if not slightly better than) the performance of a token with defocusing nonlinearity (green boxes).  
By contrast, in the case of focusing nonlinearity (blue boxes), the recorded correlation coefficients
tend to be smaller than the corresponding coefficients for a linear token (red boxes) with the same characteristics. In particular, it is worth noting that there is negligible overlap between the blue and the red boxes in all of the cases shown in the figure, and it is only the maximum recorded values which have some overlap with the red boxes.  
Comparing Fig. \ref{figure3}(a) to \ref{figure3}(b) we find that the advantage of tokens with focusing nonlinearity is getting weaker for increasing values of ${\mathscr N}$, i.e., increasing number of spectral channels. However, this advantage tends to be restored partially if we increase the 
intensity of the input light [compare Fig. \ref{figure3}(b) to \ref{figure3}(d)].  
Comparing Fig. \ref{figure3}(a) to \ref{figure3}(c) we also find that by increasing the thickness 
of the token, while keeping all other parameters constant, the advantage of the tokens with focusing nonlinearity is more pronounced, while at the same time we find a larger spread of the estimated correlation coefficients. 
Finally, as a result of Eq. (\ref{Jnu:eq}), all of these findings do not seem to depend on the actual pulse duration $T$, but rather on the number of spectral channels ${\mathscr N}$. 


\section{Concluding remarks} 

We have investigated how Kerr nonlinearity affects the sensitivity  
of a speckle to the cloning of the involved multiple-scattering optical token.
Our results suggest that for focusing nonlinearity  
the correlation between the speckles of a reference token and its clones can be smaller than 
the ones for zero or defocusing nonlinearity. Hence, PUFs that rely on tokens with 
focusing nonlinearity are expected to be more robust against cloning, in the sense that they can distinguish easier between a reference token and its clones, with high probability. 

This advantage is getting more pronounced for tokens with spectral  correlation bandwidth equal or comparable to the bandwidth of the pulse, as well as for increasing thickness of the token, and/or  increasing peak intensity of the pulse. 
The spectral correlation bandwidth of a token decreases quadratically with its 
thickness $L$ \cite{AndVolPopKatGreGig15,MouAndDefVolKatGreGig16,BeiPutLagMos11,SmaKatGuaSil12}, 
whereas for a given pulse duration the peak intensity should be  below the 
damage threshold for the medium. For our purposes, we had to vary 
the Kerr nonlinearity $n_2$ in our simulations, while keeping all other parameters fixed, which would not have been possible if the present work had been restricted to a specific material. However, for the focusing nonlinearity, where we have the most interesting findings, the main parameters pertain to fused silica 
\cite{Boyd08,Chi-etal11,Jia-etal06}. This is a well-studied material in the literature, while it has been also discussed in the framework of PUFs \cite{Iesl16}.  
The present findings are expected also to be present for other highly scattering media typically used in connection with PUFs, such as TiO$_2$ and  ZnO  \cite{Pappu02,Goorden14,AndVolPopKatGreGig15,MouAndDefVolKatGreGig16,BeiPutLagMos11,SmaKatGuaSil12}. 

The present work may serve as a benchmark for any future work on  nonlinear PUFs. 
The identification of appropriate materials and systems for the validation of the present findings in the framework of experiments is an open question which  deserves the attention of the  community in the field.


\begin{thebibliography}{1}
 \newcommand{\enquote}[1]{``#1''}


\bibitem{Pappu02}
R. Pappu,  B. Recht, J. Taylor, and N. Gershenfeld,  
Science {\bf 297}, 2026 (2002).

\bibitem{Goorden14} 
S. A. Goorden, M. Horstmann, A. P.    Mosk,  B. \v{S}kori\'{c}, and P. W. H. Pinkse,
Optica {\bf 1}, 421 (2014). 

\bibitem{Uppu19}
Uppu, R. et al. Asymmetric cryptography with physical unclonable keys. Quantum Sci. Technol. 4, 045011 (2019).

\bibitem{Iesl16}
H. Zhang, and S. Tzortzakis,  Appl. Phys. Lett. {\bf 108}, 211107 (2016).
H. Zhang, D. Di Battista, G. Zacharakis, and  S. Tzortzakis,  {\em Appl. Phys. Lett.} {\bf 109}, 039901 (2016). 

\bibitem{NikDiaSciRep17}
G. M. Nikolopoulos and E. Diamanti, Sci. Rep.  {\bf 7}, 46047 (2017).

\bibitem{Nik18}
G. M. Nikolopoulos, Phys. Rev.  A {\bf 97}, 012324 (2018).

\bibitem{FlaNikAlbFis19}
L. Fladung, G.M. Nikolopoulos, G. Alber, M. Fischlin,  Cryptography {\bf 3}, 25 (2019). 

\bibitem{Nik21}
G. M. Nikolopoulos,  Photonics {\bf 8}, 289 (2021). 

\bibitem{Mes18}
C. Mesaritakis, M. Akriotou, A. Kapsalis, E. Grivas, C. Chaintoutis, T. Nikas and  D. Syvridis, Sci. Rep.  {\bf 8 }, 9653 (2018). 

\bibitem{GiaKamBru20}
G. Gianfelici, H. Kampermann, and D. Bruss, Phys. Rev. A {\bf 101}, 042337 (2020).

\bibitem{Wang-etal21}
P. Wang, F. Chen, D. Li, S. Sun, F. Huang, T. Zhang, Q. Li, K. Chen, Y. Wan, X. Leng, Y. Yao, Phys. Rev. App. {\bf 16}, 054025 (2021). 

\bibitem{Horstmayer13}
R. Horstmayer,  B. Judkewitz,  I. M.  Vellekoop, S. Assawaworrarit, and C. Yan, 
Sci. Rep. {\bf 3}, 3543 (2013). 

\bibitem{Buch05}
J. D. R. Buchanan,  R. P.  Cowburn, A.  Jausovec, D. Petit, P.  Seem, G. Xiong,  
D. Atkinson,  K. Fenton, D. A.  Allwood,  and  M. T. Bryan,  
Nature {\bf 436}, 475 (2005). 

\bibitem{Yeh12}
C. H. Yeh, P. Y. Sung, C. H. Kuo, and R. N. Yeh,   Opt. Express {\bf 20}, 24382-24393 (2012).

\bibitem{Nik19}
G.M. Nikolopoulos, Opt. Express {\bf 27}, 29367–29379 (2019).

\bibitem{Chow-etal21}
S. Chowdhury, A. Covic, R. Y. Acharya,  S. Dupee, F. Ganji and D. Forte, 
 J. Cryptogr. Eng. (2021). https://doi.org/10.1007/s13389-021-00255-w.

\bibitem{Goodman1} 
J. C. Dainty, {\em Laser Speckle and Related Phenomena} (Springer Verlag, 1975).

\bibitem{Ruh-etal}
U. R\"uhrmair, C. Hilgers, S.  Urban,  A. Weiersh\"auser, E. Dinter,  B. Forster, and C.  Jirauschek, 
 https://eprint.iacr.org/2013/215.

\bibitem{AndVolPopKatGreGig15}
D. Andreoli, G. Volpe, S. Popoff. O. Katz, S. Gr\'esillon, and S. Gigan, 
 Sci. Rep. {\bf 5}, 10347 (2015). 
 
\bibitem{MouAndDefVolKatGreGig16} 
M. Mounaix, D. Andreoli, H. Defienne, G. Volpe, O. Katz, S. Gr\'esillon, and S. Gigan,
Phys. Rev. Lett. {\bf 116}, 253901 (2016).

\bibitem{BeiPutLagMos11}
F. van Beijnum, E. G. van Putten, A. Lagendjik, and A. P. Mosk,  Opt. Lett. {\bf 36}, 373 (2011).

\bibitem{SmaKatGuaSil12}
E. Small, O. Katz, Y. Guan, and Y. Silberberg,  Opt. Lett. {\bf 37}, 3429 (2012).

\bibitem{FroSmaDanOulDerSil17}
H. Frostig, E. Small, A. Daniel, P. Oulevey, S. Derevyanko, and Y. Silberberg,  Optica {\bf 4}, 1073 (2017).

\bibitem{Boyd08}
R. W. Boyd, {\em Nonlinear Optics} (Academic Press 2008).

\bibitem{Oka06}
K. Okamoto, {\em  Fundamentals of Optical Waveguides} (Academic Press 2006). Chap. 7.  

\bibitem{Ross14}
S. M. Ross, {\em  Introduction to Probability and Statistics for Engineers and Scientists} (Academic Press 2014). Chap. 2.  

\bibitem{Chi-etal11}
B. Chimier, O. Ut\'eza, N. Sanner, M. Sentis, T. Itina, P. Lassonde, F. L\'egar\'e, F. Vidal, and J. C. Kieffer, Phys. Rev. B {\bf 84}, 094104 (2011).

\bibitem{Jia-etal06}
T. Q. Jia, H. X. Chen, M. Huang, F. L. Zhao, X. X. Li, S. Z. Xu, H. Y. Sun, D. H. Feng, C. B. Li, X. F. Wang, R. X. Li, Z. Z. Xu, X. K. He, and H. Kuroda,  Phys. Rev. B 73, 054105 (2006).

\end{thebibliography}
\end{document}